\documentclass{ws-ijmpa}

\begin{document}

\markboth{A. Szczurek}
{Central $\eta'$ meson production}

%
\catchline{}{}{}{}{}
%

\title{DIFFRACTIVE DOUBLE-ELASTIC \\
PRODUCTION OF $\eta$' AND $\eta_c$ \\
IN THE $p p \to p X p$ REACTION}

\author{ANTONI SZCZUREK}

\address{Institute of Nuclear Physics PAN, ul. Radzikowskiego 152\\
PL-31-342 Cracow, Poland and\\
University of Rzesz\'ow, ul. Rejtana 16\\
PL-35-959 Rzesz\'ow, Poland
}

\maketitle

\pub{Received (Day Month Year)}{Revised (Day Month Year)}

\begin{abstract}
I discuss double-diffractive (double-elastic) production of the
$\eta'$ and $\eta_c$ mesons in the $pp \to p X p$ reaction within the
formalism of unintegrated gluon distribution functions (UGDF).
The contribution of $\gamma^* \gamma^* \to \eta'$ fusion is estimated.
The distributions in the Feynman $x_F$ (or rapidity), transferred 
four-momenta squared between initial and final protons ($t_1$, $t_2$)
and azimuthal angle difference between outgoing protons ($\Phi$)
are calculated and discussed. The results are compared with the WA102
data. Predictions at higher energies are presented.
\end{abstract}

\section{Introduction}

Recently the exclusive production of $\eta'$ meson in
proton-proton collisions was intensively studied close to
its production threshold at the COSY ring at KFA J\"ulich
\cite{COSY11} and at Saclay \cite{DISTO}. Here the dominant
production mechanism is exchange of several mesons (so-called
meson exchange currents) and reaction via $S_{11}$ resonance
\cite{COSY_theory}.

I present results of a recent study \cite{SPT06}
(done in collaboration with R. Pasechnik and O. Teryaev) of the same
exclusive channel but at much larger energies ($W >$ 10 GeV).
Here diffractive mechanism may be expected to be the dominant process.
In Ref.\cite{KMV99} the Regge-inspired pomeron-pomeron fusion was
considered as the dominant mechanism of the $\eta'$ production.
Here I present results obtained in the formalism with unintegrated gluon
distribution functions. Similar formalism was used recently
to calculate cross section for exclusive Higgs boson production
\cite{KKMR,Forshaw05}.
There is a chance that the formalism used for Higgs can be tested
quantitatively for exclusive (heavy) meson production where
the corresponding cross section is expected to be much bigger.

\begin{figure}[!h]    
{\includegraphics[width=0.5\textwidth]{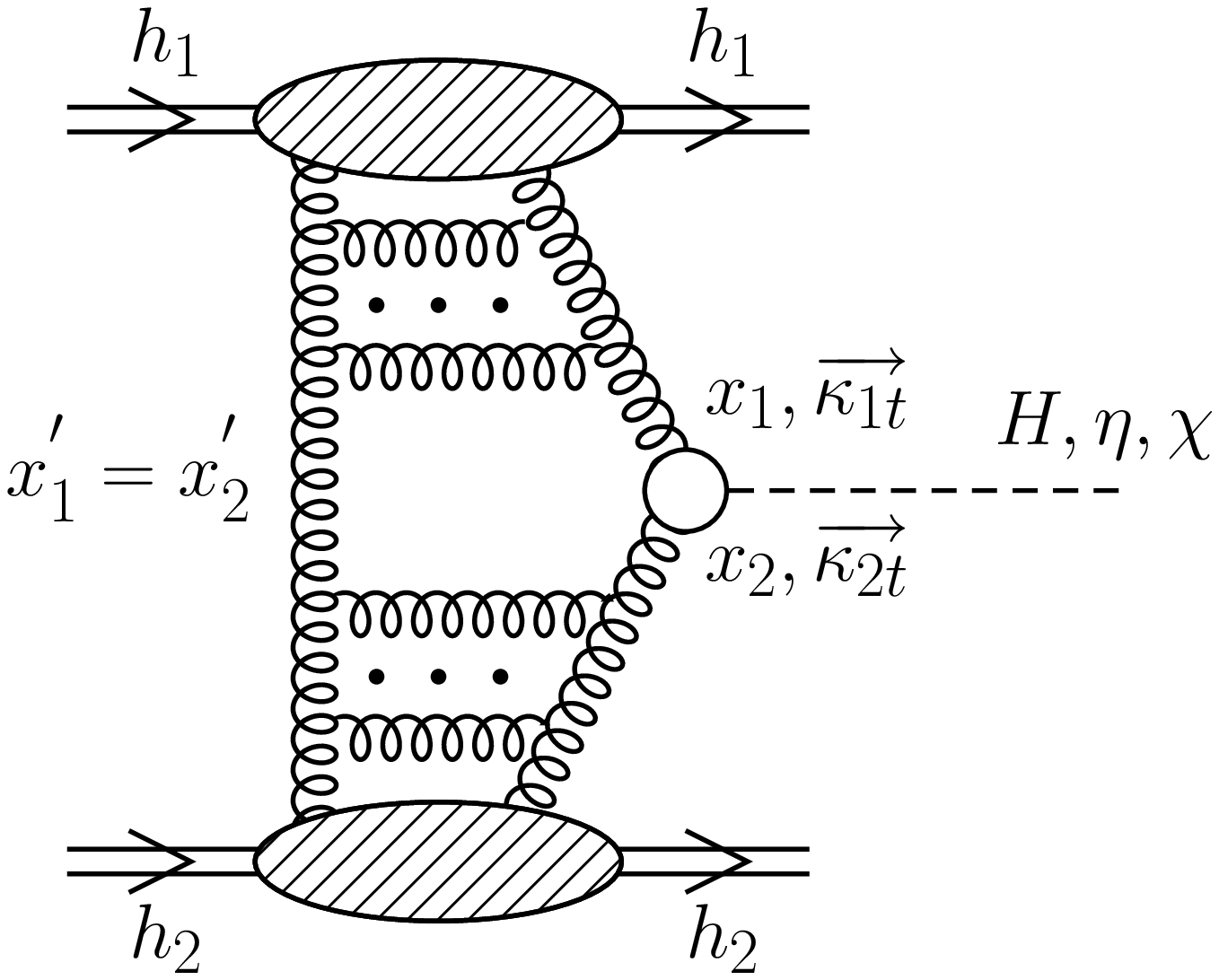}}
{\includegraphics[width=0.4\textwidth]{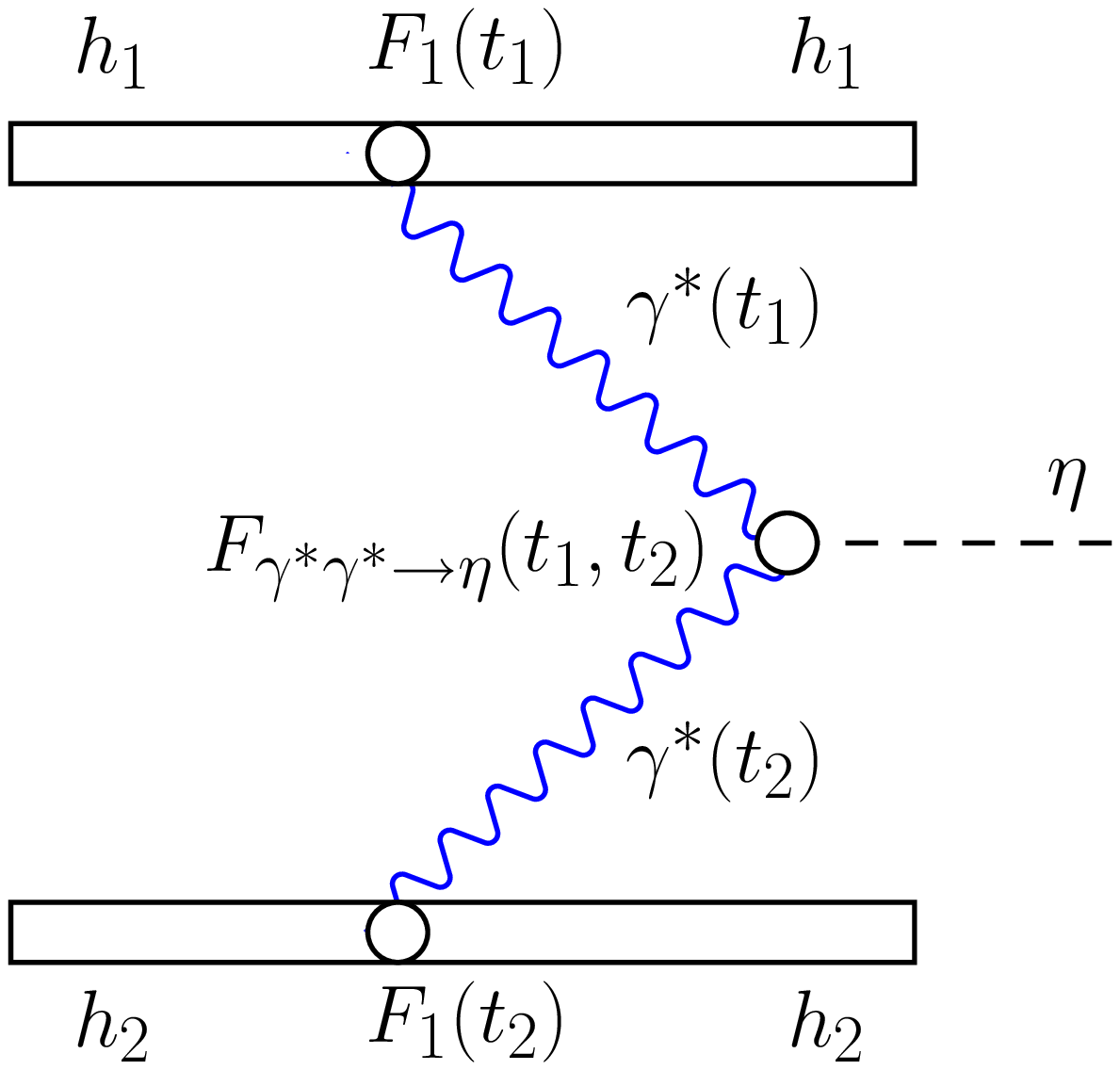}}
   \caption{\label{fig:microscopic_mechanisms}
   \small  The sketch of the bare QCD and photon-photon fusion
   mechanisms. The kinematical variables are shown in addition.}
\end{figure}

In Fig.\ref{fig:microscopic_mechanisms} I show a sketch of the QCD
mechanism of diffractive double-elastic production of $\eta'$ meson
(left diagram). For completeness, we include also photon-photon fusion
mechanism (right diagram).


\section{Formalism}

Following the formalism for the diffractive double-elastic
production of the Higgs boson developed by Kaidalov, Khoze, Martin and Ryskin
\cite{KKMR,Forshaw05} (KKMR) we write the bare QCD
amplitude for the process sketched in Fig.1 as
\begin{eqnarray}
{\cal M}_{pp \to p \eta' p}^{g^*g^*\to\eta'} =  i \, \pi^2 \int
d^2 k_{0,t} V(k_1, k_2, P_M) \frac{
f^{off}_{g,1}(x_1,x_1',k_{0,t}^2,k_{1,t}^2,t_1)
       f^{off}_{g,2}(x_2,x_2',k_{0,t}^2,k_{2,t}^2,t_2) }
{ k_{0,t}^2\, k_{1,t}^2\, k_{2,t}^2 } \, . \label{main_formula}
\end{eqnarray}
The bare amplitude above is subjected to absorption corrections which
depend on collision energy.
The vertex function $V(k_1,k_2,P_M)$ in the expression
(\ref{main_formula}) describes the coupling of two virtual gluons
to the pseudoscalar meson.
The details concerning the function $V(k_1, k_2, P_M)$ can be found
in \cite{SPT06}.

The objects $f_{g,1}^{off}(x_1,x_1',k_{0,t}^2,k_{1,t}^2,t_1)$ and
$f_{g,2}^{off}(x_2,x_2',k_{0,t}^2,k_{2,t}^2,t_2)$ appearing in
formula (\ref{main_formula}) are skewed (or off-diagonal)
unintegrated gluon distributions. They are
non-diagonal both in $x$ and $k_t^2$ space. Usual off-diagonal
gluon distributions are non-diagonal only in $x$. In the limit
$x_{1,2} \to x_{1,2}'$, $ k_{0,t}^2 \to k_{1/2,t}^2$ and $t_{1,2}
\to 0$ they become usual UGDFs.
In the general case we do not know off-diagonal UGDFs very well.
It seems reasonable, at least in the first approximation, to take
\begin{eqnarray}
f_{g,1}^{off}(x_1,x_1',k_{0,t}^2,k_{1,t}^2,t_1) &=&
\sqrt{f_{g}^{(1)}(x_1',k_{0,t}^2) \cdot
f_{g}^{(1)}(x_1,k_{1,t}^2)} \cdot F_1(t_1)
\, , \\
f_{g,2}^{off}(x_2,x_2',k_{0,t}^2,k_{2,t}^2,t_2) &=&
\sqrt{f_{g}^{(2)}(x_2',k_{0,t}^2) \cdot
f_{g}^{(2)}(x_2,k_{2,t}^2)} \cdot F_1(t_2) \, ,
\label{skewed_UGDFs}
\end{eqnarray}
where $F_1(t_1)$ and $F_1(t_2)$ are usual Dirac isoscalar nucleon
form factors and $t_1$ and $t_2$ are total four-momentum transfers
in the first and second proton line, respectively. The above
prescription is a bit arbitrary. 
It provides, however, an interpolation between different $x$ and
$k_t$ values.

Neglecting spin-flipping contributions the average matrix element
squared for the $p(\gamma^*) p(\gamma^*) \to p p \eta'$ process can
be written as \cite{SPT06}  
\begin{eqnarray}
\overline{|{\cal M}_{pp \to p\eta'p}^{\gamma^* \gamma^* \to\,
\eta'}|^2} \approx 4 s^2 e^8  \frac{F_1^2(t_1)}{t_1^2}
\frac{F_1^2(t_2)}{t_2^2} |F_{\gamma^* \gamma^*\to\,
\eta'}(k_1^2,k_2^2)|^2\, |{\bf k}_{1,t}|^2 |{\bf k}_{2,t}|^2
\sin^2(\Phi) \, . 
\label{gamma_gamma_amplitude_squared}
\end{eqnarray}
%
\section{Results}

We have shown in Ref.\cite{SPT06} that it is very difficult to describe
the only exsisting high-energy (W $\sim$ 30 GeV) data measured by
the WA102 collaboration \cite{WA102} in terms of the unintegrated
gluon distributions.
First of all, rather large cross section has been measured
experimentally. Using prescription (\ref{skewed_UGDFs}) and 
on-diagonal UGDFs from the literature we get much smaller
cross sections.
Secondly, the calculated dependence on the azimuthal
angle between the outgoing protons is highly distorted from
the $\sin^2 \Phi$ distribution, whereas the measured one is almost
a perfect $\sin^2 \Phi$ \cite{SPT06}. This signals that a rather
different mechanism plays the dominant role at this energy.
 
\begin{figure}[!h]       
 \centerline{\includegraphics[width=0.60\textwidth]{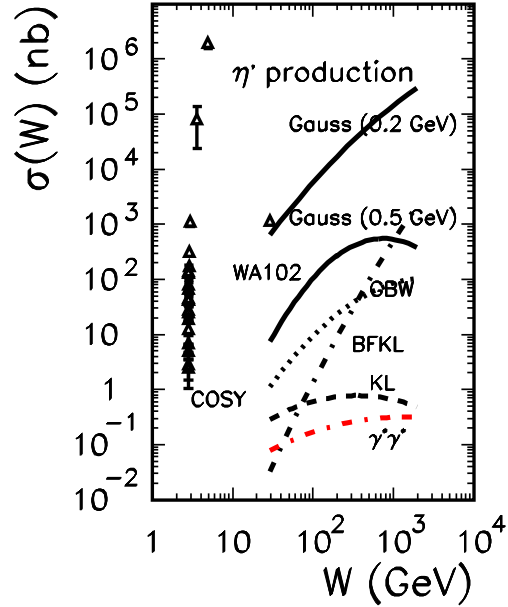}}
   \caption{ \label{fig:sig_tot_w}
\small  $\sigma_{tot}$ as a function of center of mass energy
for different UGDFs.
The $\gamma^* \gamma^*$ fusion contribution is shown by the dash-dotted
(red) line. The world experimental data are shown for reference.}
\end{figure}

In Fig.\ref{fig:sig_tot_w} I show energy dependence of
the total (integrated over kinematical variables) cross section for
the exclusive reaction $p p \to p \eta' p$ for different UGDFs.
Quite different results are obtained for different UGDFs.
This demonstrates huge sensitivity to the choice of UGDF.
The cross section with the Kharzeev-Levin type distribution (based
on the idea of gluon saturation) gives
the cross section which is small and almost idependent of beam energy.
In contrast, the BFKL distribution leads to strong energy dependence.
The sensitivity to the transverse momenta of initial gluons
can be seen by comparison of the two solid lines calculated with
the Gaussian UGDF with different smearing parameter
$\sigma_0$ = 0.2 and 0.5 GeV.
The contribution of the $\gamma^* \gamma^*$ fusion mechanism 
(red dash-dotted line) is fairly small and only slowly energy dependent.
While the QED contribution can be reliably calculated, the QCD
contribution cannot be at present fully controlled.

\begin{figure}[!hp]       
\begin{minipage}{0.49\textwidth}
\epsfxsize=\textwidth\epsfbox{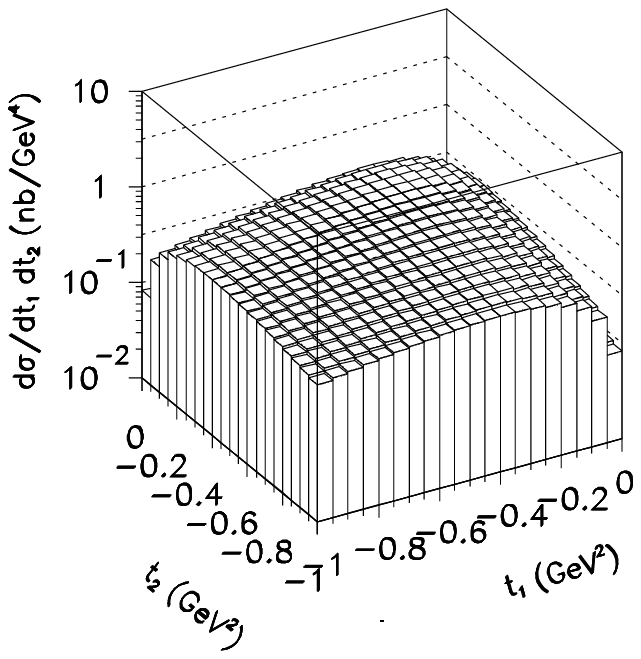}
\end{minipage}
\begin{minipage}{0.49\textwidth}
\epsfxsize=\textwidth \epsfbox{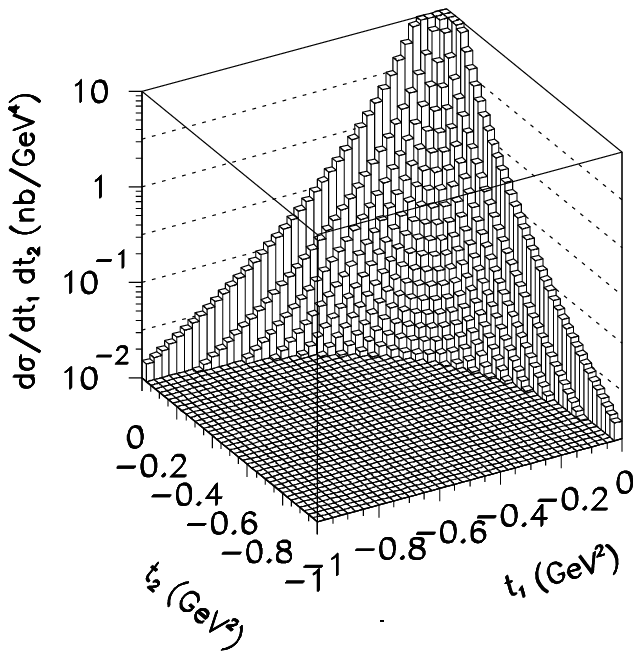}
\end{minipage}
\caption{\small Two-dimensional distribution in $t_1 \times t_2$
for the diffractive QCD mechanism (left panel), calculated with the KL
UGDF, and the $\gamma^* \gamma^*$ fusion (right panel) at
the Tevatron energy W = 1960 GeV.}
\label{fig:map_t1t2}
\end{figure}

In Fig.\ref{fig:map_t1t2} I present two-dimensional maps
$t_1 \times t_2$ of the cross section for the QCD mechanism (KL UGDF)
and the QED mechanism (Dirac terms only) for the Tevatron energy W = 1960 GeV.
If $ |t_1|, |t_2| > $ 0.5 GeV$^2$ the QED mechanism is clearly negligible.
However, at $|t_1|, |t_2| < $ 0.2 GeV$^2$ the QED mechanism may become
equally important or even dominant. In addition, it may interfere with
the QCD mechanism.

In Table~1 I have collected cross sections (in nb) for
$\eta'$ and $\eta_c$ mesons for W = 1960 GeV,
integrated over broad range of kinematical variables specified in
the table caption.
The cross sections for $\eta_c$ are very similar to corresponding
cross sections for $\eta'$ production and in some cases even bigger.


\begin{table}
\caption{\label{tab:numbers}
Comparison of the cross section (in nb) for $\eta'$
and $\eta_c$ production at Tevatron (W = 1960 GeV)
for different UGDFs.
The integration is over -4 $< y <$ 4 and -1 GeV $< t_{1,2} <$ 0.
No absorption corrections were included.}
\begin{center}
\begin{tabular}{|c|c|c|}
\hline
UGDF & $\eta'$ & $\eta_c$ \\ 
\hline
 KL                 & 0.4858(+0)       & 0.7392(+0) \\
 GBW                & 0.1034(+3)       & 0.2039(+3) \\       
 BFKL               & 0.2188(+4)       & 0.1618(+4) \\
 Gauss (0.2)        & 0.2964(+6)       & 0.3519(+8) \\
 Gauss (0.5)        & 0.3793(+3)       & 0.4417(+6) \\
\hline
$\gamma^* \gamma^*$ & 0.3095(+0)       & 0.4493(+0) \\
\hline
\end{tabular}
\end{center}
\end{table}


\section{Conclusions}

The existing models of UGDFs predict cross section much smaller
than the one obtained by the WA102 collaboration at the center-of-mass
energy W = 29.1 GeV. This may signal presence of subleading reggeons
at this ``low'' energy.

Due to a nonlocality of the loop integral our model leads to sizeable
deviations from the $\sin^2 \Phi$ dependence (predicted in the models
of one-step fusion of two vector objects).

The diffractive QCD mechanism and the photon-photon fusion lead
to quite different pattern in the $(t_1,t_2)$ space. 

Finally we have presented results for exclusive double elastic
$\eta_c$ production. Similar cross sections as for $\eta'$ production
were obtained. Also in this case the results depend strongly on
the choice of UGDF.

Measurements of the reaction(s) in the title would
help to limit or even pin down the UGDFs in the nonperturbative
region of small gluon transverse momenta where these objects
cannot be obtained as a solution of any perturbative evolution equation,
but must be rather modelled.



\end{document}